\documentclass[prd,showpacs,preprintnumbers,nofootinbib,aps]{revtex4}
\usepackage{subeqnarray}
\usepackage{epsfig,amsmath,amssymb}
\usepackage[usenames,dvipsnames]{color}
\usepackage[pagebackref=false, colorlinks=false]{hyperref}
\definecolor{redish}{rgb}{0.7,0.2,0.0}  
\definecolor{bluish}{rgb}{0.2,0.5,0.8}
\hypersetup{linkcolor=redish,          
                  citecolor=blue,        
                  filecolor=magenta,      
                  urlcolor=bluish}          
\newcommand{\ba}{\nopagebreak[3]\begin{eqnarray}}
\newcommand{\ea}{\end{eqnarray}}
\newcommand{\bii}{\begin{itemize}}
\newcommand{\eii}{\end{itemize}}

\begin{document}

\title{ Partial Inhibition of Spontaneous Voltage Recovery (`Memory Effect') of Commercial EDLC Supercapacitors by Stationary Magnetic Field}
\author{Sekhar Banerjee}%
 \email{solarsekhar@gmail.com}
\affiliation{Jatin Das Road, Kolkata 700029, India}%
\author{Parthasarathi Majumdar}
\email{partha@rkmvu.ac.in}
\affiliation{Ramakrishna Mission Vivekananda University, Belur Math 711202, India}%
\begin{abstract}

We report preliminary findings on partial inhibition of spontaneous voltage recovery under open circuit conditions, following full charge and discharge to null potential (`memory effect'), of commercial EDLC supercapacitors of Faradaic values ranging between $1-25$ Farad (rated at $2.7$ volts), upon exposure to an approximately homogeneous, stationary magnetic field of  about $100$ milliTesla.    

\end{abstract}

\maketitle
\section{Introduction}

Electric double-layer (EDL) supercapacitors of rated capacitance values ranging from $0.3 - 3,000$ Farads (at voltages $2.7-16$ volts) are now commercially available for a plethora of low voltage electronic device applications, including solar energy devices. These supercapacitors stand poised to replace batteries, at least for low voltage applications, because of their superior power-density, more efficient and more durable charge-discharge cycle and also the relative absence of toxic materials in their constituents. With increasing demands for a new kind of energy accumulator which is environmentally friendly, research on EDLCs have been proliferating over the last decade. 

One curious phenomenon exhibited by EDLCs is that of spontaneous voltage recovery, also called the `memory effect' \cite{con,lew}, occurring concomitantly with the phenomenon of self-discharge. A reasonably high faradaic value EDLC, upon charging to almost full rated potential, and then discharged to null voltage, is often seen to recover spontaneously, under open circuit conditions, a small fraction of its rated voltage. There does not appear to be a complete theoretical understanding of this observed phenomenon, although attempts have been made to analyze this based on charge redistribution models of electrochemical supercapacitors proposed by Conway et. al. \cite{con}. In these models, the existence of pores invariably occurring in the carbon electrodes, through which ions diffuse when a potential is applied across the terminals of the EDLC, leads to a redistribution of electric charge some of which can inhabit the pores. Spontaneous voltage recovery has been attributed to such a redistribution of charge, when the EDLC is no longer electrically connected to a battery source. There has been overwhelming observational evidence \cite{lew} that such a voltage recovery does indeed happen in laboratory prototypes based on novel materials \cite{and1,lew} which have been extensively researched.

Commercially available EDL supercapacitors, now manufactured by several companies internationally like Tecate, Nescap and Maxwell among others, have also been observed to exhibit the memory effect, even though information on the precise composition of the chemicals used as `ionogel' in these supercapacitors has remained a closely guarded secret. In this paper, we report observations of spontaneous voltage recovery in commercial supercapacitors, with capacitances of $1~,~10$ and $25$ Farads, as a function of time following full charging and subsequent discharging to (almost) null voltage, under open circuit conditions. The capacitors are then exposed to an approximately uniform, stationary magnetic field of $80-100$ milli-Tesla (the range arising from errors in our magnetometer measurements) for the entire period of measurement, after the completion of a full charge-discharge cycle. We observe, for {\it all} EDL supercapacitors tested, that there is an inevitable {\it inhibition} of the spontaneously recovered voltage, as a function of time : the voltage does not rise, within the same interval of time, to the value observed in absence of the magnetic field, but to a distinctly lower value. The deficiency in the spontaneously recovered voltage appears roughly to increase with increasing rated Faradaic value of the EDLC. We believe that this observation of a nontrivial effect of an applied magnetic field on the memory effect of commercially available EDLCs has been made for the first time. The inherent basic scientific interest in this observation stems from the fact that models of charge redistribution, based on tenets of electrochemistry where capacitors are circuit elements primarily using chemical kinetics and electro{\it statics}, will now have to enlarge their domain of validity, to include electro{\it dynamic} effects.
 
In the next section, we report on the details of our experiment and the devices used, with a discussion of sources of possible systematic errors. In section 3, we present our results, which show unambiguously the partial arresting of the spontaneous voltage recovery, when the external magnetic field is present. We end in section 4 with a brief summary of our results and how theoretical models currently in vogue may need to be modified substantively in order to understand our observations.

\section {Experimental details}

The devices selected for the purpose of the experiment are EDLC supercapacitors manufactured by the Tecate Group, San Diego, CA, USA, and marketed as `Powerburst ultracapacitors'. Models TPL-1.0/8X12F, TPL-10/10X30F, TPL-25/16X26F of respectively $1$, $10$ and $25$ Farad rated capacitance values at $2.7$ volts are chosen for the experiment. Initially, the capacitors are kept away from possible sources of magnetic fields, so that the effect of turning on a magnetic field on spontaneous potential recovery can be studied accurately.

Each EDLC is first charged to about $2.5$ volts (i.e., slightly lower than the rated voltage, in order to control any voltage recovery due to overcharging) by means of a $6$ volt battery eliminator. It is then discharged through a resistive load (a toy motor, of resistance of about $12$ Ohms) all the way down to $10$ millivolts. Voltage measurements are done using a standard laboratory digital voltmeter, which is certainly accurate at least down to the $10$ millivolt scale. Immediately upon discharge, the EDLC is first kept in isolation, away from any magnetic sources, under open circuit conditions. The voltage across its terminals is measured at five-minute intervals for a total of $25$ minutes. It is then polarized again by charging to $2.5$ volts and discharged through the resistive load, as before, down to $10$ millivolts. Immediately upon discharge, it is exposed to a spatially uniform, static magnetic field of about $80-100$ millitesla, and the potential across its terminals is measured at five-minute intervals, for $25$ minutes, precisely as before. This process is repeated for a pair of EDLCs of {\it each} capacitance value in our collection. 

In a subsequent experiment, the entire process is repeated for the same set of EDLCs, but voltage measurements are made every $20$ minutes over a two hour duration, both without and with exposure to the magnetic field. 

For the magnetic field source, we used the same standard ring magnet (commercially used in speakers of audio systems); its magnetic field profile is measured by a laboratory magnetometer of German make. The range of magnetic field strengths quoted, viz., $80-100$ millitesla, is really an artifact of the systematic errors in the magnetometer measurement. Since, the primary aim of the experiment is to demonstrate the effect of a magnetic field on the potential recovery, any nontrivial magnetic field suffices. Studying the variation of that effect for different magnetic field strengths is a project we have postponed to the future.  

All experiments were conducted under a controlled thermal environment of $28$ deg. C at the MSc laboratory of the Ramakrishna Mission Vivekananda University, so as to discount any major noise due to thermal fluctuations.           

\section{Results}

Spontaneous voltage recovery profiles as a function of time are plotted on a linear graph, without and with the supercapacitor exposed to the magnetic field, for one supercapacitor of each capacitance value, namely $1~,~10~,~25$ Farads. For each capacitance value, the graphs with and without magnetic field are quite distinct, with the potential profile with magnetic field always being lower in magnitude than the profile without the field. The experiment has been repeated with an additional supercapacitor of each Faradaic denomination, but with almost identical results. Also, there does not appear to be any sensitivity of the voltage recovery profile to the polarity of the magnetic field in any of the cases examined, a result that appears to be an added mystery. 

The $25$ minute data presented in Fig.s 1-3 (left panel) clearly exhibit an enhancement of the inhibitory mechanism due to magnetic field with increasing rated capacitance value, which the $2$ hour data (right panel) partially corroborate. The other aspect of interest is the recovery voltage profile as a function of time : increasing the frequency of measurement does not change the overall profile ! The change of slope appears to display a behaviour that seems to be invariant under change of time scale. 

\newpage

\includegraphics[width=7cm,height=5cm]{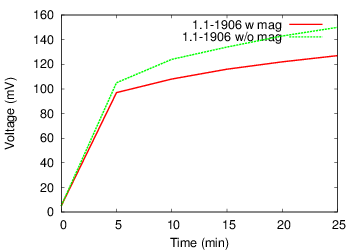}
\hglue 1cm
\includegraphics[width=7cm,height=5cm]{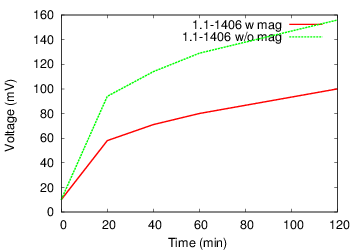} 
\vglue .5cm
\centerline{\small {\bf Fig.1} Voltage recovery  for $1$F EDLC : left - $25$ minute measurements, right -  $2$ hour measurements}
\vglue 1.5cm
\includegraphics[width=7cm,height=5cm]{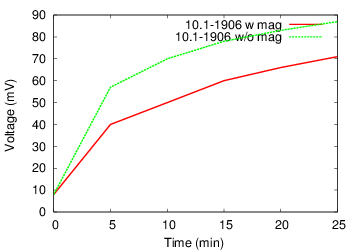}
\hglue 1cm
\includegraphics[width=7cm,height=5cm]{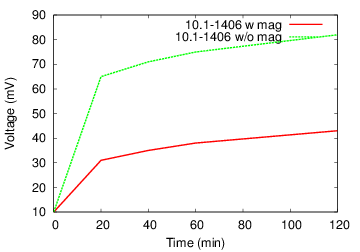} 
\vglue .5cm
\centerline{\small {\bf Fig.2} Voltage recovery for $10$F EDLC : left - $25$ minute measurements, right - $2$ hour measurements}
\vglue 1.5cm       
\includegraphics[width=7cm,height=5cm]{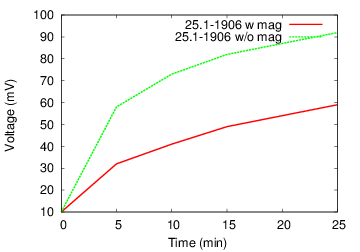}
\hglue 1cm
\includegraphics[width=7cm,height=5cm]{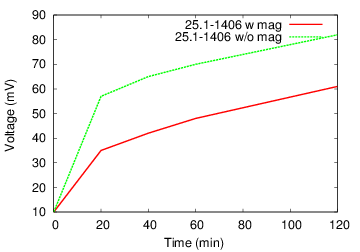} 
\vglue .5cm
\centerline{\small {\bf Fig.3} Voltage recovery for $25$F EDLC : left - $25$ minute measurements, right - $2$ hour measurements}

\newpage
\section{Summary and Discussion}

Our primary experimental findings are :
\begin{itemize}
\item Commercial EDLCs of Faradaic values $1~,~10$ and $25$ F (rated, at $2.7$ volts) are subjected to a complete charge-discharge cycle each. Their rates of open-circuit spontaneous voltage recovery is then studied. Next, after another charge-discharge cycle, the supercapacitors are then exposed to a stationary, approximately uniform magnetic field of about $100$ milliTesla, and their open-circuit spontaneous voltage recovery rates are studied again. It is observed that there is a marked decrease in the rates in the presence of the magnetic field.
\item The decrease seems to be an increasing function of the rated capacitance of the EDLCs; thus the effect is most pronounced for the $25$ F EDLC.
\item This inhibitory phenomenon is apparently insensitive to the polarity of the applied magnetic field.
\item The experiment has been performed with a single kind of magnet, and therefore no data can be reported on the variations with respect to changes in the magnetic field.
\item However, in a very preliminary experiment performed alongside the ones reported above, a spatially varying magnetic field appears to have an enhancing effect on the rate of open-circuit spontaneous voltage recovery. This result is of course subject to further verification and observations.
\end{itemize}
   
At the present moment we do not have a complete theoretical understanding of this phenomenon. Despite the existence of a good bit of literature on the effect of magnetic fields on electrochemical processes in general \cite{isr}, we have found no analysis in the literature on the effect of magnetic fields on EDLCs, especially of the commercial variety, and that too on the phenomenon of open-circuit spontaneous voltage recovery. This, thus, seems to be a new direction of research where much remains to be done. Even if we regard spontaneous voltage recovery as a kind of a dual effect of self-discharge of an EDLC which has indeed been studied substantively by Conway et. al. \cite{con}, Andreas et. al. \cite{andr} and more recently reviewed comprehensively by Lewandowski et. al. \cite{lew}, it is by no means obvious how the mechanisms for self-discharge discussed in these works are to be modified to take into account the effect of the external magnetic field, in order to explain our observations. In ref. \cite{isr}, a hydrodynamic model on the effect of magnetic fields on general electrochemical phenomena purports to explain it in terms of a convection type of magnetohydrodynamic process, rather than in terms of Lorentz force interactions on individual ionic charges. It is not clear to us that this is valid for the carbon nanotube EDLCs under consideration. Also, we have not yet been able to discern whether the predictions of this model will eventually lead to the kind of inhibitory phenomenon observed by us for the open-circuit spontaneous voltage recovery. In our opinion, there is room for enormous progress towards a complete understanding of the observations made so far.

In concluding, we refer to the work of Andreas et. al. \cite{andr} on the effect of an iron contaminant present in the electrodes; the authors claim that their observations are explained by Conway's overcharge-induced activation mechanism for the positive electrode, and Conway's diffusion-based mechanism for the negative electrode. We have not been able to extrapolate from this work towards any further understanding of the effect of magnetic fields. Of course, there also more needs to be done, in terms of studying the effect of spatially and temporally varying magnetic fields on open-circuit voltage recovery. We hope to report on these in the near future. 

\noindent {\bf Acknowledgement} We thank Kaliprasanna Mondal for help with the laboratory magnetometer for measuring the magnetic field profile of our ring magnets, and for discussions.

\end{document}